\documentclass[prl,twocolumn,nofootinbib]{revtex4}
\usepackage{amsmath,amssymb}
\usepackage{graphicx}
\def\beq{\begin{equation}}
\def\eeq{\end{equation}}
\def\bea{\begin{eqnarray}}
\def\eea{\end{eqnarray}}

\begin{document}
\vspace{6.0cm}
\title{Supersymmetric Adjoint SU(5)}
\vspace{0.3cm}
\author{Pavel Fileviez P\'erez}
\vspace{0.3cm}
\email{fileviez@cftp.ist.utl.pt}
\vspace{0.3cm}
\affiliation{
Centro de F{\'\i}sica Te\'orica de Part{\'\i}culas \\
Departamento de F{\'\i}sica.\ Instituto Superior T\'ecnico \\
Avenida Rovisco Pais 1, 1049-001 Lisboa, Portugal}
%%%%%%%%%%%%%%%%%%%%%%%%%%%%%%%%%%%%%%%%%%%%%%%%%%%%%%%%%%%%%%%%%%%%
\begin{abstract}
Recently we have proposed a renormalizable grand unified theory, 
based on the $SU(5)$ gauge symmetry, where the neutrino masses are 
generated through the type I and type III seesaw mechanisms. 
In this letter we study the supersymmetric version of this theory. 
As in the non-susy version it is possible to generate all fermion masses 
with the minimal number of Higgses, the theory predicts one massless 
neutrino and the leptogenesis mechanism can be realized. All contributions 
to the decay of the proton and the properties of neutralinos are discussed.
This theory can be considered as the simplest renormalizable supersymmetric 
grand unified theory based on the $SU(5)$ gauge symmetry since it has the 
minimal number of superfields and free parameters.
\end{abstract}
%%%%%%%%%%%%%%%%%%%%%%%%%%%%%%%%%%%%%%%%%%%%%%%%%%%%%%%%%%%%%%%%%%%%
%\pacs{}
\maketitle
%%%%%%%%%%%%%%%%%%%%%%%%%%%%%%%%%%%%%%%%%%%%%%%%%%%%%%%%%%%%%%%%%%%%
%Introduction.
%%%%%%%%%%%%%%%%%%%%%%%%%%%%%%%%%%%%%%%%%%%%%%%%%%%%%%%%%%%%%%%%%%%

\section{ I. Introduction} 

The so-called Grand Unified Theories (GUTs) can be considered as one 
of the most appealing candidates for physics beyond the Standard Model (SM). 
These theories predict the unification of the electromagnetic, weak and 
strong interactions at the high scale, $M_{GUT} \approx 10^{14-16}$ GeV, 
the quantization of the electric charge, the value of $\sin^2 \theta_W (M_{GUT}) = 3/8$ 
at the GUT scale, the decay of the proton and the existence of vector and 
scalar leptoquarks. The simplest grand unified theory was proposed in 
reference~\cite{GG}. This theory is based on $SU(5)$ gauge symmetry 
and one Standard Model family is partially unified in the 
anti-fundamental ${\bf \overline{5}}$ and antisymmetric ${\bf 10}$ representations. 
The Higgs sector is composed of ${\bf {{24}}_H}$ and ${\bf {{5}}_H}$, while the GUT 
symmetry is broken down to the Standard Model by the vacuum expectation 
value of the Higgs singlet field in ${\bf {24}_H}$, and the SM Higgs resides in ${\bf {{5}}_H}$. 
As is well-known this theory is ruled out since in this case the unification of 
gauge couplings is in disagreement with the values of 
$\alpha_s$, $\sin \theta_W$ and $\alpha_{em}$ at the electroweak scale. Recently, 
several efforts has been made in order to define the simplest realistic extension 
of the Georgi-Glashow model. See references~\cite{Ilja-Pavel},~\cite{Borut-Goran} 
and~\cite{Adjoint} for different minimal realistic non-supersymmetric grand unified theories.   

The minimal supersymmetric $SU(5)$ theory was discussed 
for the first time in reference~\cite{SUSYSU(5)}. In this 
case one generation of matter of the Minimal Supersymmetric 
Standard Model (MSSM) is unified in two chiral superfields 
$\mathbf{\hat{\bar 5}} = (\hat{d}^C,\hat{L})$ and 
$\mathbf{\hat{10}} = (\hat{u}^C,\hat{Q},\hat{e}^C)$, 
while the Higgs sector is composed of $\mathbf{\hat 5_H}=(\hat{T}, 
\hat{H}_1)$, $\mathbf{\hat{\bar 5}_H}=(\hat{\overline{T}},\hat{\overline{H}}_1)$, 
and $\mathbf{\hat{24}_H}$. In our notation the SM decomposition 
of the adjoint Higgs superfield reads as $\mathbf{\hat{24}_H}=
(\hat{\Sigma}_8, \hat{\Sigma}_3, \hat{\Sigma}_{(3,2)}, 
\hat{\Sigma}_{(\bar{3}, 2)}, \hat{\Sigma}_{24})=(8,1,0)
\bigoplus (1,3,0) \bigoplus (3,2,-5/6)\bigoplus (\overline{3},2,5/6) 
\bigoplus (1,1,0)$. As is well-known the renormalizable version 
of this theory is ruled out since the relation between $Y_E$ 
and $Y_D$, $Y_E=Y_D^T$, is in disagreement with the experimental 
values of the fermion masses at the low scale and the neutrinos 
are massless if the so-called R-parity is conserved. 
See reference~\cite{SUSY-IPG} for the most general constraints 
coming from unification and~\cite{BPG} for all possible dimension 
five contributions to the decay of the proton in this context. 

In this letter we write down and study the supersymmetric version 
of the theory proposed in reference~\cite{Adjoint}. In this 
renormalizable theory the fermion masses are generated with the 
minimal set of Higgs bosons, $\mathbf{5_H}$ and $\mathbf{45_H}$. 
The neutrino masses are generated through the type I~\cite{TypeI} 
and type III~\cite{TypeIII} seesaw mechanisms using the 
fermionic ${\mathbf{24}}$ representation. We conclude that 
if we want R-parity as a symmetry of the theory we have to 
introduce one matter chiral superfield $\mathbf{\hat{24}}$.  
We refer to this model as ``Supersymmetric Adjoint SU(5)''. 
As in the non-supersymmetric version the theory predicts 
one massless neutrino and the leptogenesis mechanism 
can be realized. We discuss the LLLL and RRRR contributions 
to proton decay and the properties of neutralinos. 
This theory can be considered as the simplest renormalizable 
grand unified theory based on $SU(5)$ since it has the minimal 
number of chiral superfields and free parameters.      
         
%%%%%%%%%%%%%%%%%%%%%%%%%%%%%%%%%%%%%%%%%%%%%%%%%%%%%%%%%%%%%%%%%%%%%%%%%%%%%%%%%%%%%%%%
\section{II. SUSY Adjoint $SU(5)$} 
%%%%%%%%%%%%%%%%%%%%%%%%%%%%%%%%%%%%%%%%%%%%%%%%%%%%%%%%%%%%%%%%%%%%%%%%%%%%%%%%%%%%%%%%%

Recently, a realistic renormalizable grand unified theory based 
on $SU(5)$ gauge symmetry has been proposed, where the Higgs sector 
is composed of $\mathbf{5_H}$, $\mathbf{24_H}$, and $\mathbf{45_H}$~\cite{Adjoint}. 
In this case an extra matter multiplet in the adjoint representation has 
been added in order to generate neutrino masses through the type I and type III 
seesaw mechanisms. This model is consistent with all constraints coming 
from proton decay, predicts one massless neutrino at tree level, and 
the leptogenesis mechanism can be realized~\cite{Adjoint}. 
Let us discuss in this section the supersymmetric version of this model.     

As is well-known in the minimal supersymmetric $SU(5)$~\cite{SUSYSU(5)} the 
MSSM chiral superfields are unified in $\mathbf{\hat{\bar 5}}$ and 
$\mathbf{\hat{10}}$, while its Higgs sector comprises $\mathbf{\hat 5_H}$, 
$\mathbf{\hat{\bar 5}_H}$, and $\mathbf{\hat{24}_H}$. Now, in order to write 
down the supersymmetric version of the realistic grand unified theory 
proposed in reference~\cite{Adjoint} we have to introduce three extra chiral superfields, 
$\mathbf{\hat{45}_H}$, $\mathbf{\hat{\overline{45}}_H}$ and $\mathbf{\hat{24}}$. 
Therefore, our Higgs sector will be composed of $\mathbf{{\hat{\overline{5}}}_H}$, 
$\mathbf{\hat{{{5}}}_H}$, $\mathbf{\hat{24}_H}$, $\mathbf{\hat{45}_H} =( \hat{\Phi}_1, \hat{\Phi}_2, \hat{\Phi}_3, \hat{\Phi}_4, \hat{\Phi}_5, \hat{\Phi}_6, \hat{H}_2)=({8},{2},1/2)\bigoplus$$(\overline{{6}},{1},-1/3)$ $\bigoplus({3},{3},-1/3)\bigoplus(\overline{{3}},{2},-7/6)
\bigoplus({3},{1},-1/3)\bigoplus$$(\overline{{3}},{1},4/3)$ 
$\bigoplus({1}, {2},1/2)$, and $\mathbf{\hat{\overline{45}}_H} =( \hat{\overline \Phi}_1, \hat{\overline \Phi}_2, \hat{\overline \Phi}_3, \hat{\overline \Phi}_4, \hat{\overline \Phi}_5, 
\hat{\overline \Phi}_6, \hat{\overline H}_2)=({8},{2},- 1/2)\bigoplus 
({{6}},{1},1/3)\bigoplus({\overline{3}},{3},1/3) \bigoplus({{3}},{2},7/6)
\bigoplus $ $(\overline{3},{1},1/3)\bigoplus$$({{3}},{1},- 4/3)\bigoplus({1}, {2},- 1/2)$. The fields in the ${45}$ representation satisfy the following conditions: 
$({45})^{\alpha \beta}_{\delta} = - ({45})^{\beta \alpha}_{\delta}$, 
$\sum_{\alpha=1}^5 ({45})^{\alpha \beta}_{\alpha} = 0$, $v_{45} = \langle 45_H \rangle^{1 5}_{1}= \langle 45_H \rangle^{2 5}_{2}= \langle 45_H \rangle^{3 5}_{3}$, and $v_{\overline{45}} = \langle \overline{45}_H \rangle_{1 5}^{1}= \langle \overline{45}_H \rangle_{2 5}^{2} = \langle \overline{45}_H \rangle_{3 5}^{3}$. 

In this model the Yukawa superpotential for charged fermions reads as:
\begin{eqnarray}
{\cal{W}}_{0} &=& {\hat{10}} \ \hat{\overline{{5}}} \ \left( Y_1 \ \hat{\overline{5}}_H 
\ + \  Y_2 \ \hat{\overline{45}}_H \right) \ + \nonumber\\
&+& {\hat{10}} \ {\hat{10}} \ \left( Y_3 \ {\hat{5}_H} \ + \ Y_4 \ \hat{45}_H \right)  
\end{eqnarray}
and the relation between the masses for charged leptons and down quarks is given by:
\begin{eqnarray}
M_D \ - \ M_E^T &=& 8 \ Y_2 \ v_{\overline{45}}  \label{GJ}
\end{eqnarray}
where $Y_2$ is an arbitrary $3 \times 3$ matrix. As is well-known 
the relation between the masses of $\tau$ lepton and $b$ quark, 
$m_b (M_{GUT})= m_{\tau} (M_{GUT})$, is in agreement with the 
experiment. Therefore, the $Y_2$ matrix only must modify the 
relation between the masses of quarks and leptons of the first and 
second generation. See reference~\cite{Ross-Serna} for a recent 
study of the relation between fermion masses in supersymmetric 
scenarios.      

There are three different possibilities to generate the neutrino 
masses at tree level in the context of SUSY SU(5) models: 

\begin{itemize}

\item We can add at least two fermionic superfields for the singlets 
and generate neutrino masses through the type I seesaw~\cite{TypeI} mechanism.

\item We can add two Higgs chiral superfields $\mathbf{\hat{15}_H}$ and 
$\mathbf{\hat{\overline{15}}_H}$ to generate neutrino masses through 
type II seesaw~\cite{TypeII} mechanism.

\item One can generate neutrino masses through the type III~\cite{TypeIII} 
and type I seesaw mechanisms adding just one fermionic $\mathbf{\hat {24}}$ 
chiral superfield. The last possibility has been realized at the 
renormalizable level in the model proposed in reference~\cite{Adjoint}. 
Therefore, in order to realize the mechanism a new chiral supermultiplet: 
$
\mathbf{\hat{24}}= (\hat{\rho}_8,\hat{\rho}_3, \hat{\rho}_{(3,2)}, \hat{\rho}_{(\bar{3}, 2)},
\hat{\rho}_{0})=({8},{1},0)\bigoplus({1},{3},0)\bigoplus({3},{2},-5/6)
\bigoplus(\overline{{3}},{2},5/6)
$
$
\bigoplus({1},{1},0) 
$ 
has to be introduced~\cite{Ma}. 
\end{itemize}

The idea of using extra matter in the adjoint 
representation to generate neutrino masses through the type I and type III 
seesaw mechanisms was pointed out for the first time in reference~\cite{Ma} 
in the context of SUSY $SU(5)$ and in reference~\cite{Borut-Goran} 
in the context of non-SUSY $SU(5)$. It is important to say that this possibility 
is very appealing since we have to introduce only one extra chiral matter 
superfield and there is no need to introduce $SU(5)$ singlets. 

Since in this letter we are interested in the supersymmetric version 
of the model proposed in reference~\cite{Adjoint} a new matter chiral 
superfield has to be introduced only if we want to have the so-called 
matter parity as a symmetry of the theory. Matter parity is defined 
as $M=(-1)^{3(B-L)}=(-1)^{2S} R$, where $M=-1$ for all matter 
superfields and $M=1$ for the Higgses and gauge superfields. In the case 
that matter parity is not conserved the neutrino masses can be generated 
through the M-parity violating interactions $\epsilon_i \hat{\bar{5}}_i \hat{5}_H$ 
and $\eta_i \hat{\bar{5}}_i \hat{24}_H \hat{5}_H$. Particularly, in the second 
term we have an $SU(2)$ fermionic triplet needed for type III seesaw mechanism. 
In this letter we want to keep matter-parity as a symmetry of the theory 
to avoid the dimension four contributions to the decay of proton coming 
from $\lambda_{ijk} \hat{10}_i \hat{\bar{5}}_j \hat{\bar{5}}_k$ and have 
the lightest neutralino as a good candidate for the cold dark matter of 
the universe. 

The new superpotential relevant for neutrino masses in this context 
is given by:
\begin{equation}
{\cal W}_{1} = c_i \ \hat{\overline{5}}_i \ \hat{24} \ \hat{5}_H \ 
+ \ p_i \ \hat{\overline{5}}_i \ \hat{24} \ \hat{45}_H   
\end{equation}      
Notice from Eq.~(1) and Eq.~(3) the possibility to generate all fermion 
masses, including the neutrino masses, with the Higgs chiral superfields 
$\mathbf{\hat{5}_H}$, $\mathbf{\hat{\bar{5}}_H}$, 
$\mathbf{\hat{{45}}_H}$ and $\mathbf{\hat{\overline{45}}_H}$. 
As in the non-susy model the Higgses in the $\mathbf{45}$ representation 
play a crucial role to generate masses for charged fermions 
and neutrinos as well.  

There are also new relevant interactions between $\mathbf{\hat{24}}$ 
and $\mathbf{\hat{24}_H}$ in this model:
\begin{eqnarray}
{\cal W}_{2} &=& m_{\Sigma} \ \text{Tr} \ \hat{24}_H^2 \ + \ 
{\lambda_{\Sigma}} \  \text{Tr} \ \hat{24}_H^3 \ + \  
m \ \text{Tr} \ \hat{24}^2 \nonumber \\
\ & + & \ \lambda \ \text{Tr} \ (\hat{24}^2 \hat{24}_H)  
\end{eqnarray}
Notice that there are only two extra terms since matter parity is conserved.
Once $24_H$ gets the expectation value, 
$\langle 24_H \rangle = 2 m_{\Sigma} \ diag (2,2,2, -3, -3) / 3 \lambda_{\Sigma}$, 
the masses of the fields living in $24$ are given by: 
\begin{eqnarray}
M_{\rho_0} &=& m - \frac{2 m_{\Sigma} \ \lambda}{3 \lambda_{\Sigma}},\\
M_{\rho_3} &=& m -\frac{2 \lambda \ m_{\Sigma}}{\lambda_{\Sigma}},\\
M_{\rho_8} &=& m + \frac{4 \lambda \ m_{\Sigma}}{3 \lambda_{\Sigma}},
\end{eqnarray}
and
\begin{eqnarray}
M_{\rho_{(3,2)}} &=& M_{\rho_{(\bar{3},2)}} = m -\frac{\lambda \ m_{\Sigma}}{3 \lambda_{\Sigma}}.
\end{eqnarray}
From the above equations we can see that when the fermionic 
triplet $\rho_3$, responsible for type III seesaw mechanism, 
is very light the rest of the fields living in $\mathbf{\hat{24}}$ 
have to be heavy if we do not assume a very small value 
for the $\lambda$ parameter. The GUT symmetry is broken as 
usual and $\mathbf{\hat{24}}$ does not get expectation value. 

Since our Higgs sector is composed of $\mathbf{\hat{5}_H}$, 
$\mathbf{\hat{\bar{5}}_H}$, $\mathbf{\hat{{45}}_H}$, 
$\mathbf{\hat{\overline{45}}_H}$ and $\mathbf{\hat{24}_H}$ there are 
also additional interactions between the different Higgs 
chiral superfields in the theory:

\begin{eqnarray}
{\cal W}_{3} &=& m_{H} \ \hat{\overline{5}}_H  \hat{5}_H \ + \ 
\lambda_H \ \hat{\overline{5}}_H  \hat{24}_H \hat{5}_H \nonumber \\
\ & + & \ c_H \ \hat{\overline{5}}_H  \hat{24}_H \hat{45}_H \ + \ 
b_H \ \hat{\overline{45}}_H  \hat{24}_H \hat{5}_H \nonumber \\    
\ & + & \  m_{45} \ \hat{\overline{45}}_H  \hat{45}_H \ + \ 
a_H \ \hat{\overline{45}}_H  \hat{45}_H \hat{24}_H 
\end{eqnarray} 

Notice the simplicity of the model. Unfortunately, 
the scalar sector of the non-supersymmetric grand unified 
theory proposed in reference~\cite{Adjoint} is not very 
simple since there are many possible interactions 
between ${\bf 5_H}$, ${\bf 24_H}$ and ${\bf 45_H}$. We have the same 
problem in any renormalizable non-supersymmetric 
grand unified model. Supersymmetric Adjoint $SU(5)$, 
the model proposed in this letter, can be considered 
as the simplest supersymmetric grand unified theory 
based on $SU(5)$ since it has the minimal number 
of chiral superfields and free parameters. 

%%%%%%%%%%%%%%%%%%%%%%%%%%%%%%%%%%%%%%%%%%%%%%%%%%%%%%%%%%%
\section{III. Phenomenological Aspects: {\small proton decay, neutrino 
masses, neutralinos and leptogenesis}} 
%%%%%%%%%%%%%%%%%%%%%%%%%%%%%%%%%%%%%%%%%%%%%%%%%%%%%%%%%%%

In this section we will discuss the most relevant 
phenomenological and cosmological aspects of this 
proposal. However, the detailed analysis of those 
issues is beyond the scope of this letter. As is well known 
the most important prediction coming from the unification 
of fundamental forces is proton decay. See reference~\cite{review} 
for a review and~\cite{experiments} for future proton decay 
experiments. In this model there are several multiplets that 
mediate proton decay. We have the usual gauge $d=6$ contributions, 
mediated by the superheavy gauge bosons 
$V=({3},{2},-5/6)\bigoplus(\overline{{3}},{2},5/6)$, and 
Higgs $d=6$ contributions mediated by the fields $T$,$\overline{T}$, $\Phi_3$, 
$\overline{\Phi}_3$, $\Phi_5$, $\overline{\Phi}_5$, $\Phi_6$, 
and $\overline{\Phi}_6$. The most important contributions to the 
decay of the proton in supersymmetric scenarios are the dimension five 
contributions if the so-called matter parity is conserved. In our model 
the most important proton decay contributions are mediated by the 
superpartners of the above fields: $\tilde{T}$, $\tilde{\overline{T}}$, 
$\tilde{\Phi}_3$, $\tilde{\overline{\Phi}}_3$, $\tilde{\Phi}_5$, 
$\tilde{\overline{\Phi}}_5$, $\tilde{\Phi}_6$, and 
$\tilde{\overline{\Phi}}_6$. Let us discuss the different 
LLLL and RRRR contributions. The so-called LLLL 
effective operators, $\hat{Q} \ \hat{Q} \ \hat{Q} \ \hat{L}$, 
are generated once we integrate out the fields 
$\tilde{T}$, $\tilde{\overline{T}}$, $\tilde{\Phi}_3$, $\tilde{\overline{\Phi}}_3$, 
$\tilde{\Phi}_5$, and $\tilde{\overline{\Phi}}_5$. 
The RRRR contributions, $\hat{U}^C \ \hat{E}^C \ \hat{U}^C \ \hat{D}^C$, 
are due to the presence of the fields $\tilde{T}$, $\tilde{\overline{T}}$,   
$\tilde{\Phi}_5$, $\tilde{\overline{\Phi}}_5$, $\tilde{\Phi}_6$, and 
$\tilde{\overline{\Phi}}_6$. As is well-known those fields 
have to be very heavy in order to satisfy the experimental 
bounds on the proton decay lifetime. There are also new contributions 
to nucleon decay in this context. Once we compute the F-terms of 
the fields in the adjoint representation we find new dimension 
five contributions. However, these contributions are suppressed 
since they are proportional to $m_W / M_T^2$~\cite{new}.

Let us analyze how we could suppress the LLLL and RRRR 
contributions mentioned above. The different dimension five 
contributions are obtained through the mixings between  
${\bf \hat{5}_H}$ and ${\bf \hat{\bar{5}}_H}$ (the usual contributions 
in minimal $SU(5)$), ${\bf \hat{5}_H}$ and ${\bf \hat{\overline{45}}_H}$ 
 (proportional to $b_H$), ${\bf \hat{\bar{5}}_H}$ and ${\bf \hat{45}_H}$ 
 (proportional to $Y_4$), and through the mixing between 
${\bf \hat{{45}}_H}$ and ${\bf \hat{\overline{45}}_H}$ 
(proportional to $Y_4$). Notice that in the case when $Y_4$ 
and $b_H$ are very small the only relevant LLLL and RRRR 
contributions to the decay of the proton are due to 
the mixing between ${\bf \hat{5}_H}$ and ${\bf \hat{\bar{5}}_H}$ 
since all others are suppressed. Now, without assuming 
large masses for sfermions one can satisfy the 
experimental bounds on the proton decay lifetime 
if the unification scale and the mass of the 
triplets $\tilde{T}$ and $\tilde{\overline{T}}$ 
is around $10^{17}$ GeV. In this model it is easy to 
realize this scenario. The complete numerical analysis 
of the proton decay issue in this model is beyond 
the scope of this letter and will be studied 
in detail in a future publication~\cite{new}. 

As in the non-supersymmetric version of the model~\cite{Adjoint}, 
integrating out the singlet $\rho_0$ and the neutral component 
of the $SU(2)$ fermionic triplet $\rho_3$ in $\mathbf{24}$, 
the neutrino mass matrix reads as

\begin{eqnarray}
M^\nu_{ij} & = & \frac{a_i a_j}{M_{\rho_3}} \ + \ \frac{b_i b_j}{M_{\rho_0}}, 
\end{eqnarray} 
with
\begin{equation}
a_i = c_i  v_5 \ - \  3 p_i v_{45},
\end{equation}
and
\begin{equation}
b_i = \frac{\sqrt{15}}{2} \left( \frac{c_i  v_5}{5} \ + \ p_i v_{45}  \right). 
\end{equation}   
The theory predicts one massless neutrino at tree level, 
this is one of the main predictions.
Therefore, we could have a \textit{normal neutrino mass 
hierarchy}: $m_1=0$, $m_2=\sqrt{\Delta m_{sun}^2}$ and 
$m_3=\sqrt{\Delta m_{sun}^2 + \Delta m_{atm}^2}$ or \textit{the 
inverted neutrino mass hierarchy}: $m_3=0$, $m_2=\sqrt{\Delta m_{atm}^2}$ 
and $m_1=\sqrt{\Delta m_{atm}^2 - \Delta m_{sun}^2}$. 
$\Delta m_{sun}^2 \approx 8 \times 10^{-5}$ eV$^{2}$ 
and $\Delta m_{atm}^2 \approx 2.5 \times 10^{-3}$ eV$^{2}$ are the 
mass-squared differences of solar and atmospheric neutrino 
oscillations, respectively. The Higgs sector is composed of 
two pairs of Higgs chiral superfields , $\hat{H}_1$, $\hat{\overline{H}}_1$, 
$\hat{H}_2$ and $\hat{\overline{H}}_2$. See reference~\cite{Manuel} 
for phenomenological aspects of supersymmetric models with 
several chiral Higgs superfields. 

In this model the neutralino states are defined as 
$\widetilde{\Psi}^0_i = N_{i1} \ \widetilde{B} \ + \ N_{i2} \ \widetilde{W}_3^0 \ 
+ \ N_{i3} \ \widetilde{\overline{H}_1^0} \ + \ N_{i4} \ \widetilde{{H}_1^0} \ + \ N_{i5} 
\ \widetilde{\overline{H}_{2}^0} \ + \ N_{i6} \ \widetilde{{H}_{2}^0}$ and 
the mass matrix for them reads as:

\begin{equation}
\left(\begin{array}{cccccc}
M_1 & 0 & -\frac{1}{2} g' v_{\bar 5} & \frac{1}{2} g' v_5  & -\frac{1}{2} g' v_{\overline{45}} &  
\frac{1}{2} g' v_{45} 
\\
0 & M_2 & \frac{1}{2} g v_{\bar 5} & -\frac{1}{2} g v_5  & \frac{1}{2} g v_{\overline{45}} 
& -\frac{1}{2} g v_{45}  
\\
-\frac{1}{2} g' v_{\bar 5} & \frac{1}{2} g v_{\bar 5} & 0 & -\mu_1 & 0 & -\mu_2 
\\
\frac{1}{2} g' v_5 & -\frac{1}{2} g v_5 & -\mu_1 & 0 & -\mu_3 &  0 
\\
- \frac{1}{2} g' v_{\overline{45}} & \frac{1}{2} g v_{\overline{45}} & 0 & -\mu_3 &  0 & -\mu_4 
\\
\frac{1}{2} g' v_{45} & -\frac{1}{2} g v_{45} & -\mu_2 & 0 & -\mu_4 &  0 
\\
\\
\end{array}\right)
\end{equation} 
where $v_5 = \langle 5_H \rangle$, and $v_{\bar 5} = \langle \bar{5}_H \rangle$.
The $\mu_i$ parameters in the above matrix are given by:
${\mu_1} = - m_H \ + \ 2 m_{\Sigma} \ \lambda_H / \lambda_{\Sigma}$,
${\mu_2} = - 10 \ c_H \ m_{\Sigma} / \lambda_{\Sigma}$,
${\mu_3} = 10 \ b_H \ m_{\Sigma} / \lambda_{\Sigma}$, 
and ${\mu_4} = 12 \ m_{45} \ + \ 22 \ a_H \ m_{\Sigma} / \lambda_{\Sigma}$.
At low energy we have just one pair of light Higgsinos 
with mass $\mu_{eff}$. 

Also as in the non-supersymmetric version of the model 
it is possible to realize the leptogenesis mechanism 
in this context (For a review see~\cite{leptogenesis}.). 
In this case a net B-L asymmetry can be generated through 
the  out of equilibrium decays of the fields $\rho_0$ and $\rho_3$ 
and their superpartners in the adjoint representation. 

Let us now compare our model with the unrealistic minimal 
renormalizable SUSY $SU(5)$. In our model we have three extra 
chiral superfields, two Higgs chiral superfields 
$\mathbf{\hat{45}_H}$ and $\mathbf{\hat{{\overline{45}}}_H}$, 
and one matter chiral superfield $\mathbf{\hat{24}}$. 
All those fields could modify the predictions coming 
from the unification of gauge couplings. In $\mathbf{\hat{24}}$ 
we have four superfields, $\hat{\rho}_8$, $\hat{\rho}_3$, 
$\hat{\rho}_{(3,2)}$, and $\hat{\rho}_{(\bar{3}, 2)}$, which 
contribute to the running of gauge couplings. However, 
only $\hat{\rho}_3$ could help us to improve the unification 
in agreement with the values of $\alpha_s(M_Z)$, $\alpha_{em} (M_Z)$ 
and $\sin \theta_W (M_Z)$ since it has positive 
(negative) contribution to $b_2 - b_3$ ($b_1 - b_2$). 
Here $b_i$ stands for the different beta functions.
Notice that in the limit when $\lambda \to 0$, see Eq.~(4), 
the mass splitting between the fields in the adjoint 
representation is very small, they do not modify the 
running of the gauge couplings at one-loop level 
and still we can generate mass for two neutrinos.
In the case of the $\mathbf{\hat{45}_H}$ and 
$\mathbf{\hat{\overline{45}}_H}$ there are 
four fields, $\hat{\Phi}_3$, $\hat{\overline{\Phi}}_3$, 
$\hat{H}_2$ and $\hat{\overline{H}}_2$ with positive 
(negative) contributions to $b_2 - b_3$ ($b_1 - b_2$).
However, as we have discussed above the fields 
in $\hat{\Phi}_3$ and $\hat{\overline{\Phi}}_3$ mediate 
proton decay, therefore they have to be at the GUT scale 
if we do not suppress their contributions. A detailed 
numerical analysis of the unification of gauge couplings 
in this model is beyond the scope of this letter. All the 
phenomenological and cosmological aspects mentioned 
above will be studied in detail in a future 
publication~\cite{new}.  

%%%%%%%%%%%%%%%%%%%%%%%%%%%%%%%%%%%%%%%%
\section{IV. Summary and Outlook}
%%%%%%%%%%%%%%%%%%%%%%%%%%%%%%%%%%%%%%%%

In this letter we have written and studied the minimal 
supersymmetric version of the renormalizable grand 
unified theory based on the $SU(5)$ gauge symmetry 
with extra matter in the adjoint representation 
proposed in reference~\cite{Adjoint}. We refer to this 
model as ``Supersymmetric Adjoint $SU(5)$''. As in the 
non-susy version of the theory it is possible to generate 
all fermion masses, including the neutrino masses, with 
the minimal number of Higgses. The theory predicts one 
massless neutrino at tree level and a net $B-L$ asymmetry 
can be generated in the early universe through the 
out of equilibrium decays of the fermions responsible 
for type I and type III seesaw mechanisms and 
their superpartners in the adjoint matter chiral superfield.
The LLLL and RRRR contributions to proton decay and 
the properties of neutralinos have been discussed. 
A detailed analysis of the predictions for neutrino masses, 
the constraints coming from leptogenesis, the numerical 
analysis for proton decay and the unification of 
gauge couplings will be published in a future publication.
The theory presented in this work can be considered as the 
simplest renormalizable supersymmetric grand unified theory 
based on the $SU(5)$ gauge symmetry since it has the 
minimal number of superfields.
\\
\\
%%%%%%%%%%%%%%%%%%%%%%%%%%%%%%%%%%%%%%%%%%%%%%%%%%%%%%%%%%%%%%%%%%
\textit{Acknowledgments.}
{\small I would like to thank M. Drees, P. Nath and M. N. Rebelo 
for the careful reading of the manuscript and very useful comments. 
I thank G.~Walsch for comments on the manuscript. 
This work has been supported by { Funda\c{c}\~{a}o 
para a Ci\^{e}ncia e a Tecnologia} (FCT, Portugal) through the 
projects CFTP, POCTI-SFA-2-777, PDCT/FP/63914/2005, PDCT/FP/63912/2005 
and a fellowship under project POCTI/FNU/44409/2002. 
I would like to thank the theory division at CERN for support and 
hospitality.}
%%%%%%%%%%%%%%%%%%%%%%%%%%%%%%%%%%%%%%%%%%%%%%%%%%%%%%%%%%%%%%%%%%

%%%%%%%%%%%%%%%%%%%%%%%%%%%%%%%%%%%%%%%%%%%%%%%%%%%%%%%%%%%%%%%%%%%%%%%%

\end{document}